\documentclass[aps,prd,11pt,reprint,showpacs,floatfix,nofootinbib,superscriptaddress]{revtex4-1}
\usepackage{color,amsmath,amssymb,amsthm,graphicx,latexsym}
\usepackage{graphics,bm}
\usepackage{amssymb}
\usepackage{array}
\usepackage{mathtools}
\usepackage{psfrag}

\usepackage{amsthm}
\usepackage{newlfont}
\usepackage{esdiff}
\usepackage{multirow}
\usepackage{graphics}
\usepackage[english]{babel}
\usepackage{graphicx}
\usepackage{ae,aecompl,color}

\usepackage{bm}
\usepackage{booktabs}

\usepackage[colorlinks=true,citecolor=cyan,urlcolor=blue,bookmarks=true,bookmarks=true,bookmarksopen=true,bookmarksnumbered=true,bookmarksopenlevel=3]{hyperref}

\makeatletter
\DeclareRobustCommand*{\bfseries}{%
  \not@math@alphabet\bfseries\mathbf
  \fontseries\bfdefault\selectfont
  \boldmath
}
\makeatother

\definecolor{airforceblue}{rgb}{0.36, 0.54, 0.66}
\definecolor{steelblue}{rgb}{0.27, 0.51, 0.71}
\definecolor{amber}{rgb}{1.0, 0.49, 0.0}

\begin{document}
\newcommand{\bd}{\begin{document}}
\newcommand{\ed}{\end{document}}
\newcommand{\bc}{\begin{center}}
\newcommand{\ec}{\end{center}}
\newcommand{\bfr}{\begin{flushright}}
\newcommand{\efr}{\end{flushright}}
\newcommand{\lt}{\left}
\newcommand{\rt}{\right}
\newcommand{\vs}{\vspace}
\newcommand{\hs}{\hspace}
\newcommand{\beq}{\begin{equation}}
\newcommand{\eeq}{\end{equation}}
\newcommand{\lb}{\linebreak}
\newcommand{\pb}{\pagebreak}
\newcommand{\mb}{\makebox}
\newcommand{\fb}{\framebox}
\newcommand{\mc}{\multicolumn}
\newcommand{\ben}{\begin{enumerate}}
\newcommand{\een}{\end{enumerate}}
\newcommand{\bit}{\begin{itemize}}
\newcommand{\eit}{\end{itemize}}
\newcommand{\ovl}{\overline}
\newcommand{\un}{\underline}
\newcommand{\lefq}{\lefteqn}
\newcommand{\ba}{\begin{array}}
\newcommand{\ea}{\end{array}}
\newcommand{\beqa}{\begin{eqnarray}}
\newcommand{\eeqa}{\end{eqnarray}}
\newcommand{\beqas}{\begin{eqnarray*}}
\newcommand{\eeqas}{\end{eqnarray*}}
\newcommand{\bfg}{\begin{figure}}
\newcommand{\efg}{\end{figure}}
\newcommand{\bds}{\begin{displaymath}}
\newcommand{\eds}{\end{displaymath}}
\newcommand{\btb}{\begin{tabbing}}
\newcommand{\etb}{\end{tabbing}}
\newcommand{\para}{\parallel}
\newcommand{\pad}{\partial}
\newcommand{\nn}{\nonumber}
\newcommand{\la}{\leftarrow}
\newcommand{\ra}{\rightarrow}
\newcommand{\lgla}{\longleftarrow}
\newcommand{\lgra}{\longrightarrow}
\newcommand{\La}{\Leftarrow}\newcommand{\Ra}{\Rightarrow}
\newcommand{\Lra}{\Leftrightarrow}
\newcommand{\Lgla}{\Longleftarrow}
\newcommand{\Lgra}{\Longrightarrow}
\newcommand{\lan}{\langle}
\newcommand{\ran}{\rangle}
\renewcommand{\a}{\alpha}
\renewcommand{\b}{\beta}
\newcommand{\g}{\gamma}
\newcommand{\G}{\Gamma}
\renewcommand{\d}{\delta}
\newcommand{\eps}{\epsilon}
\newcommand{\s}{\sigma}
\newcommand{\D}{\Delta}
\newcommand{\vare}{\varepsilon}
\newcommand{\pr}{\prime}
\newcommand{\ro}{\rho}
\newcommand{\nab}{\nabla}
\newcommand{\m}{\mu}
\newcommand{\n}{\nu}
\newcommand{\Sg}{\Sigma}
\newcommand{\p}{\pi}
\newcommand{\R}{I\!\!R}
\newcommand{\om}{\omega}
\newcommand{\Om}{\Omega}
\newcommand{\ze}{\zeta}
\newcommand{\vart}{\vartheta}
\newcommand{\lam}{\lambda}
\newcommand{\tri}{\triangle}
\newcommand{\f}{\frac}
\newcommand{\iny}{\infty}
\newcommand{\pro}{\propto}
\newcommand{\np}{\newpage}

\title{Solutions of the Bogoliubov-de Gennes equation \\with  position dependent Fermi--velocity and gap profiles} 

\author{\textsc{M.~Presilla}}
\affiliation{Dipartimento di Fisica e Geologia, Universit\`a degli Studi di Perugia, Via A.~Pascoli, I-06123 Perugia, Italy}
\author{\textsc{O.~Panella}}
\affiliation{Istituto Nazionale di Fisica Nucleare, Sezione di Perugia, Via A.~Pascoli, I-06123 Perugia, Italy}
\email[({\bf Corresponding Author})\\ Email: ]{orlando.panella@pg.infn.it }

\author{\textsc{P.~Roy}}
\affiliation{Physics and Applied Mathematics Unit, Indian Statistical Institute, Kolkata-700108, India}

\date{\today}

\begin{abstract}
It is shown that bound state solutions of the one dimensional Bogoliubov-de Gennes (BdG) equation may exist when the Fermi velocity becomes dependent on the space coordinate. The existence of bound states in continuum (BIC) like solutions has also been confirmed  both in the normal phase as well as in the super-conducting phase.  We also show that a combination of Fermi velocity and  gap  parameter step-like profiles provides scattering solutions  with normal reflection and transmission.
\end{abstract}
\vspace{2cm}

\maketitle

\noindent\underline{Introduction}\\  The Bogoliubov-de Gennes (BdG) equation plays a particularly important role in the context of superconductivity~\cite{Beenakker:2006aa,Tinkham:2004aa}. This equation is also essential in the study of Andreev refelction~\cite{Andreev:1964aa,Titov:2006aa,Beenakker:2008aa}. Recently the BdG equation with a linear potential \cite{Ke:2016aa} has been studied in the presence of modified uncertainty principle or a minimal length formalism~\cite{Kempf:1994aa,Kempf:1995aa}.   Since its inception the Minimal length formalism~\cite{Garay:1995aa,Gross:1988aa} has been studied in various contexts. In particular various quantum mechanical models have been studied within the minimal length formalism to understand the effect of the minimal length parameter on observables like energy~\cite{Chang:2002aa,Dadic:2003aa,Gemba:2007aa,Fityo:2006aa,Akhoury:2003aa,Benczik:2005aa,Quesne:2005aa,Nouicer:2006aa,Quesne:2007aa,Jana:2009aa}. These phenomena actually can be used to obtain a bound on the minimal length/momentum parameter. In ref.~\cite{Ke:2016aa} it was shown that  in the BdG equation bound states do not exist in the absence of minimal length (maximal momentum) while in the presence of minimal length (maximal momentum) bound states do exist with energy depending on the  relevant parameter introduced by the model. 

In this article our objective is to suggest an alternative scenario to create bound states or bound states in continuum (BIC) in the BdG equation. Here our approach would be to allow space modulation of the Fermi velocity $v_F$. Such a scenario has been considered earlier in various contexts e.g, in graphene~\cite{Raoux:2010aa,Concha:2010aa,Krstajic:2011aa,Liu:2012aa}, one dimensional hetero-structures~\cite{Peres:2009aa} etc. It was also shown that in one dimensional hetero-structures space modulation of Fermi velocity helps creation of bound states as well as BIC~\cite{Panella:2012aa}. Here it will be shown that such a scenario may be replicated in the case of the BdG equation also. Finally we also address the question of normal reflectance and transmission within the BdG equation. It will be shown that a step-like order parameter ($\Delta$) and Fermi velocity ($v_F$) profiles entails scattering solutions with non zero reflection. \\

\noindent\underline{Formalism}\\To begin with we note that the BdG Hamiltonian in the Andreev approximation is given by \cite{Tinkham:2004aa,Andreev:1964aa,Ke:2016aa}
\beq\label{hamiltonian}
H=\left(\ba{cc} v_F\,\,p_x & \Delta \\ \Delta^{\!\!*} & -v_F\,\,p_x\ea\right)
\eeq
where $v_F, \Delta$ are respectively the Fermi velocity and the superconductor order (gap) parameter. However it may be noted that when the Fermi velocity depends on the space coordinate the operator $H$ no longer remains Hermitian and in order to maintain Hermiticity the term $v_Fp_x$ has to be replaced by $\sqrt{v_F(x)}p_x\sqrt{v_F(x)}$~\cite{Peres:2009aa}. With this replacement  and allowing for a position dependent $\Delta$ the Hamiltonian in Eq.~(\ref{hamiltonian}) becomes:
\beq\label{HvFx}
H=\left(\ba{cc} \sqrt{v_F(x)}\, p_x\sqrt{v_F(x)} & \Delta(x) \\ \Delta^{\!\!*}(x) & -\sqrt{v_F(x)}\, p_x\sqrt{v_F(x)}\ea\right)
\eeq
In the following we will discuss the solutions of the corresponding component equations associated to the BdG Hamiltonian of Eq.~\eqref{HvFx} upon introducing a two-component spinor $\psi^T =(\psi_1,\psi_2)$:
\begin{equation}
\label{timedepDirac}
i\hbar\frac{\partial}{\partial t}\left(\ba{c} \psi_1 \\ \psi_2\ea\right)=H\left(\ba{c} \psi_1 \\ \psi_2\ea\right)
\end{equation}  
dealing in particular with stationary state solutions of the associated eigenvalue equation $H\psi= E\psi$:
\begin{equation}
\label{eigenvalue}
H\left(\ba{c} \psi_1 \\ \psi_2\ea\right)=E\left(\ba{c} \psi_1 \\ \psi_2\ea\right)\, .
\end{equation}
First of all we  easily verified that the time-dependent Dirac equation, c.f. Eq.~\eqref{timedepDirac}, associated to the Hamiltonian  given in Eq.\eqref{hamiltonian} admits a conserved probability current.
The continuity equation: 
\begin{equation}
\frac{\partial }{\partial x} \,j(x,t) +\frac{\partial}{\partial t}\, \rho(x,t) =0  \, ,
\end{equation}
can be derived by using straightforward procedures, i.e. combining appropriately the two components of Eq.~\eqref{timedepDirac}. The
 probability current $j(x,t)$ is found to be given by:
\begin{equation}
\label{current}
j(x,t)= (\sqrt{v_F}\psi_1)^*(\sqrt{v_F}\psi_1)-(\sqrt{v_F}\psi_2)^*(\sqrt{v_F}\psi_2)
\end{equation}and the time dependent probability density is as usual $\rho(x,t)=\langle \psi | \psi \rangle = \psi_{1}^{*} \psi_{1}^{\phantom{*}}+ \psi_{2}^{*} \psi_{2}^{\phantom{*}}$.

In terms of the components the eigenvalue equation given in Eq.(\ref{eigenvalue}) reads
\beq
\ba{l}
\phantom{-}\sqrt{v_F(x)}\,p_x\sqrt{v_F(x)}\,\psi_1+\Delta^{\!\!\phantom{*}}(x)\,\, \psi_2=E\,\psi_1\\
-\sqrt{v_F(x)}\,p_x\sqrt{v_F(x)}\,\psi_2+\Delta^{\!\!*}(x)\,\, \psi_1=E\,\psi_2
\ea
\eeq
In order to find the components $\psi_{1,2}$ we now multiply the above equations from the left by $\sqrt{v_F(x)}$ and writing 
\beq\label{psiphi}
\sqrt{v_F(x)}\psi_{1,2}=\phi_{1,2}
\eeq
we obtain:
\begin{subequations}
\label{intert2}
\begin{align}
\label{intert2a}
\phantom{-}v_F(x)\,p_x\phi_1+\Delta^{\!\!\phantom{*}}(x)\,\phi_2&=E\,\phi_1\\
\label{intert2b}
-v_F(x)\,p_x\phi_2+\Delta^{\!\!*}(x)\,\phi_1&=E\,\phi_2
\end{align}
\end{subequations}
The components can now be easily decoupled. For example, it can be easily shown that the component $\phi_1$ satisfies the equation
\begin{equation}
\label{secondordereq}
-v^{2}_{F}(x) \frac{d^{2} \phi_{1}}{dx^{2}} -v_{F}(x) v^{\prime}_{F}(x) \frac{d \phi_{1}}{dx} +\frac{\lvert\Delta(x)\rvert^2}{\hbar^2} \phi_1= \frac{E^2}{\hbar^2}\, \phi_{1} \, .
\end{equation}
The lower component $\phi_2$ is then computed from Eq.~\eqref{intert2a}. We shall now choose specific velocity and gap profiles to examine creation of bound states,  BIC and scattering solutions.
\vspace{0.1cm}

\noindent\underline{Constant gap  and hyperbolic Fermi velocity profile.}\\  First we consider  a step-like velocity profile along with constant gap:  
\begin{subequations}
\label{v1}
\begin{align}
\label{v1a}v_F(x)&=v_0~\text{cosh}^2(\a x),~~~~v_0>0\, ,\\
\label{v1b}\Delta(x)&=\Delta =\text{const.}
\end{align}
\end{subequations}
Eq.~\eqref{secondordereq} can be further reduced and brought to the standard Schr\"odinger form :
\beq\label{phi1}
\frac{d^{2} \phi_{1}}{dq^{2}} + \epsilon^2 \phi_{1}=0,
\eeq		
where $q$ and $\eps$ are given by
	\begin{equation}	\label{q}
		q=\int \frac{dx}{v_{F}(x)},~~~~	\epsilon^2=\frac{E^2-\vert \Delta \vert ^{2}}{\hbar^{2}}
\end{equation}		
In this case we find from Eq. (\ref{q})
\beq
q=\f{\text{tanh}(\a x)}{\a v_0},~~-\f{1}{\a v_0}<q<\f{1}{\a v_0}
\eeq
We now consider $E^2>|\Delta|^2$ i.e, $\eps>0$. In this case the solution of Eq.(\ref{phi1}) is given by
\beq\label{phi1above}
\phi_1(q)=c_1\,\sin(\eps q)+c_2\,\cos(\eps q)\\
\eeq
where $c_{1,2}$ are arbitrary constants. 
For bound states the wave function $\phi_1(q)$ should vanish at the boundary values i.e, 
\beq\label{bc}
\phi_1(\pm \f{1}{\a v_0})=0
\eeq
Then from Eq.~(\ref{phi1above}) it follows that
\begin{subequations}
\begin{align}
\phi_{n1}(q)&=N_n \sin \left( \sqrt{\frac{E_n^{2}-\vert \Delta \vert^{2}}{\hbar^{2}}}~q \right)\\
E_n^{2}&=n^{2}\pi^{2}\alpha^{2}v_{0}^{2}\hbar^{2}+ \vert \Delta \vert^{2}, \qquad n=1,2,3, \dots 
\end{align}
\end{subequations}
where $N_n$ is a normalization constant. We now go back to the $x$ space to obtain $\psi_1(x)$ via Eq.(\ref{psiphi}).  Subsequently one has to use (\ref{intert2}) to obtain the component $\psi_{2}(x)$. Thus the complete solution of the BdG equation is given by:
\begin{widetext}
\beq
\label{discretelevels}
\ba{l}
\psi_n=\left(\ba{c} \psi_{n1} \\ \psi_{n2}\ea\right)={\cal N}_n\left(\ba{c} \text{sech} (\alpha x) \sin \left(n\pi \tanh (\alpha x) \right) \\ \text{sech} (\alpha x) \left[ \frac{E_n}{ \Delta } \sin \left( n\pi \tanh ( \alpha x) \right)+i {n \pi \gamma \frac{\Delta^{\!*}}{|\Delta|}} \cos \left( n \pi \tanh ( \alpha x) \right) \right]\ea\right)\\E_n=\vert \Delta \vert\sqrt{(n\pi\gamma)^2+ 1},~~\qquad n=1,2,3, \dots ~~\qquad \gamma=\frac{\alpha v_0\hbar}{|\Delta|}
\ea		
\eeq
\end{widetext}
\begin{figure*}[t]
\includegraphics[scale=0.525]{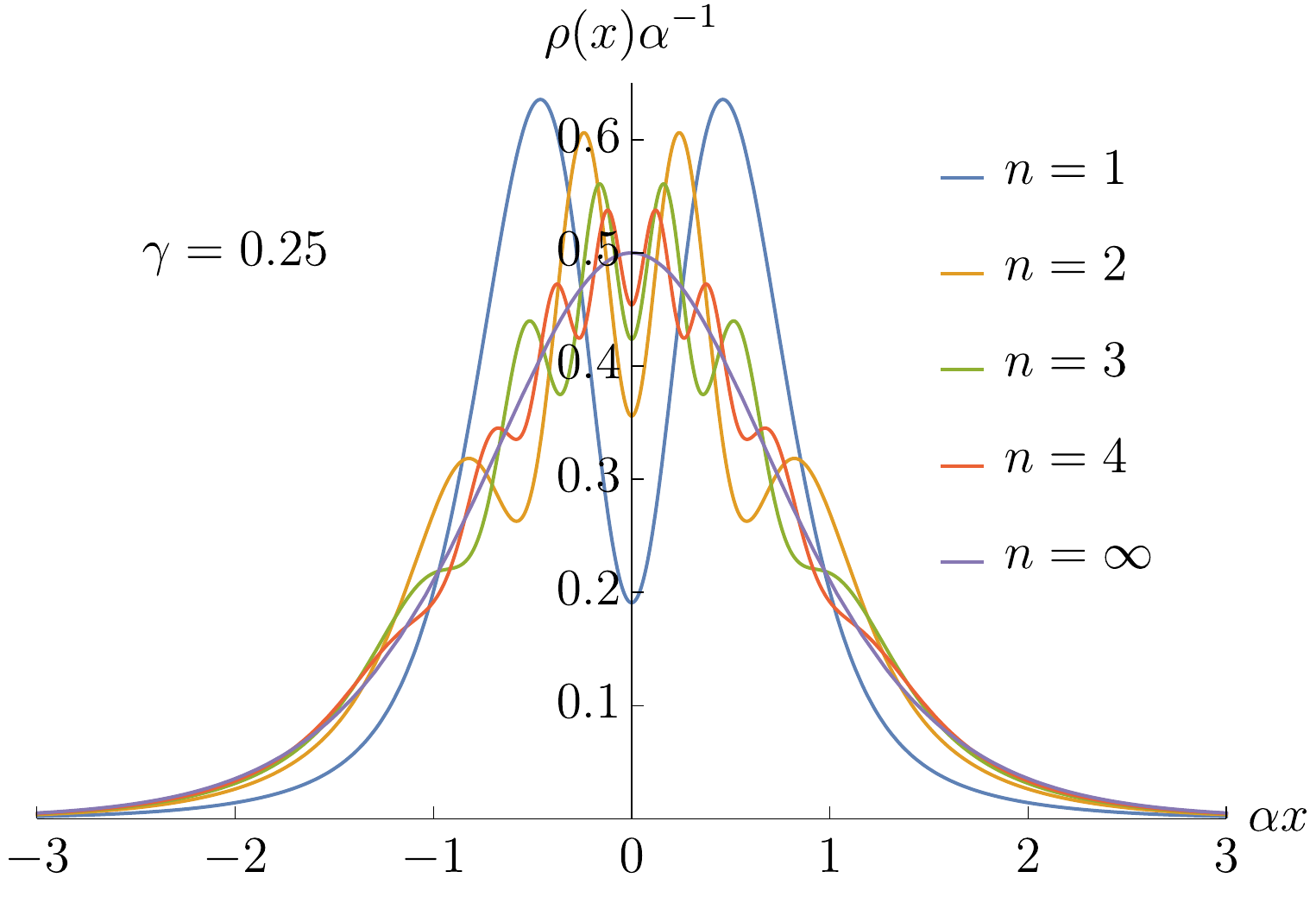}
\includegraphics[scale=0.525]{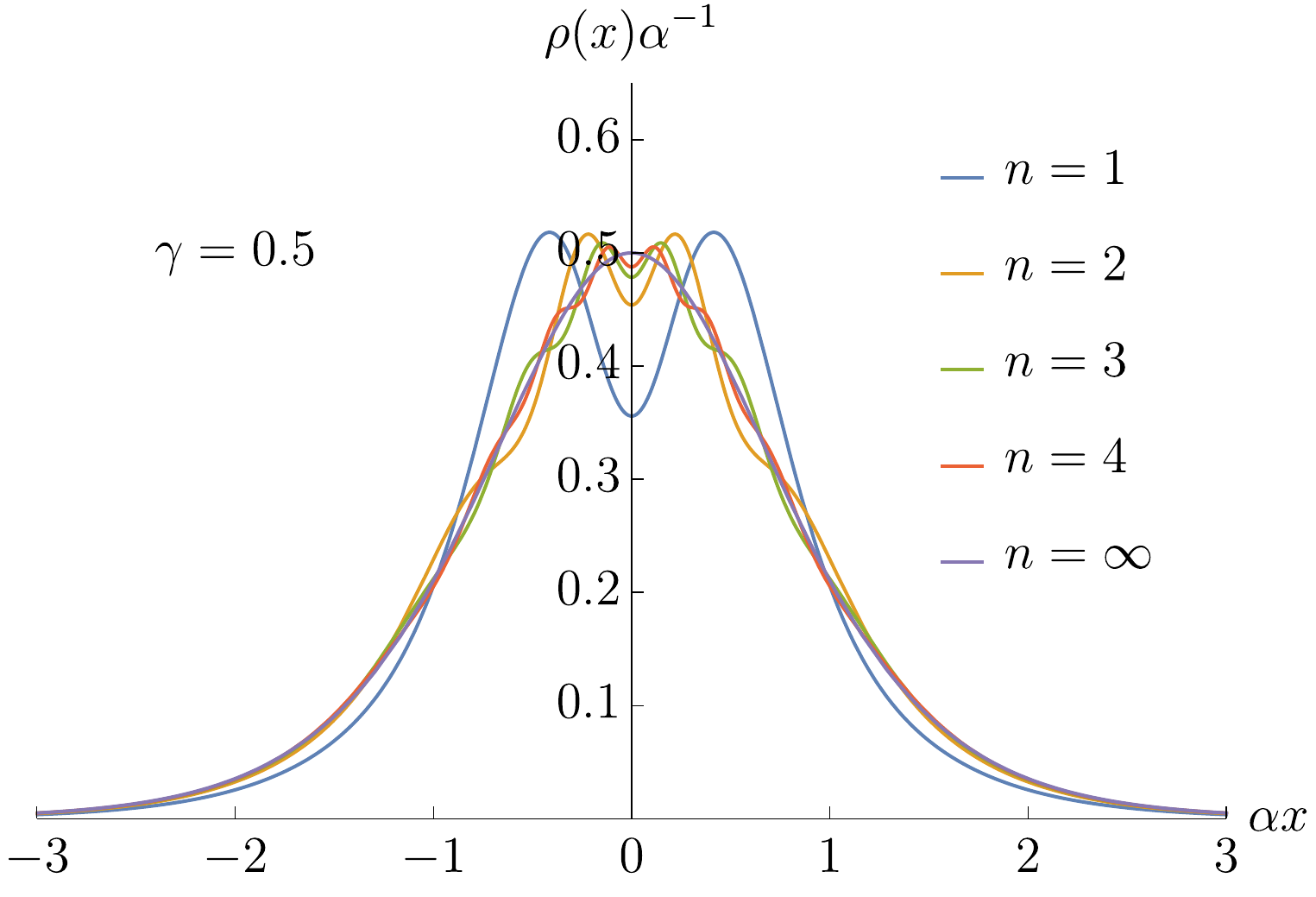}
\includegraphics[scale=0.525]{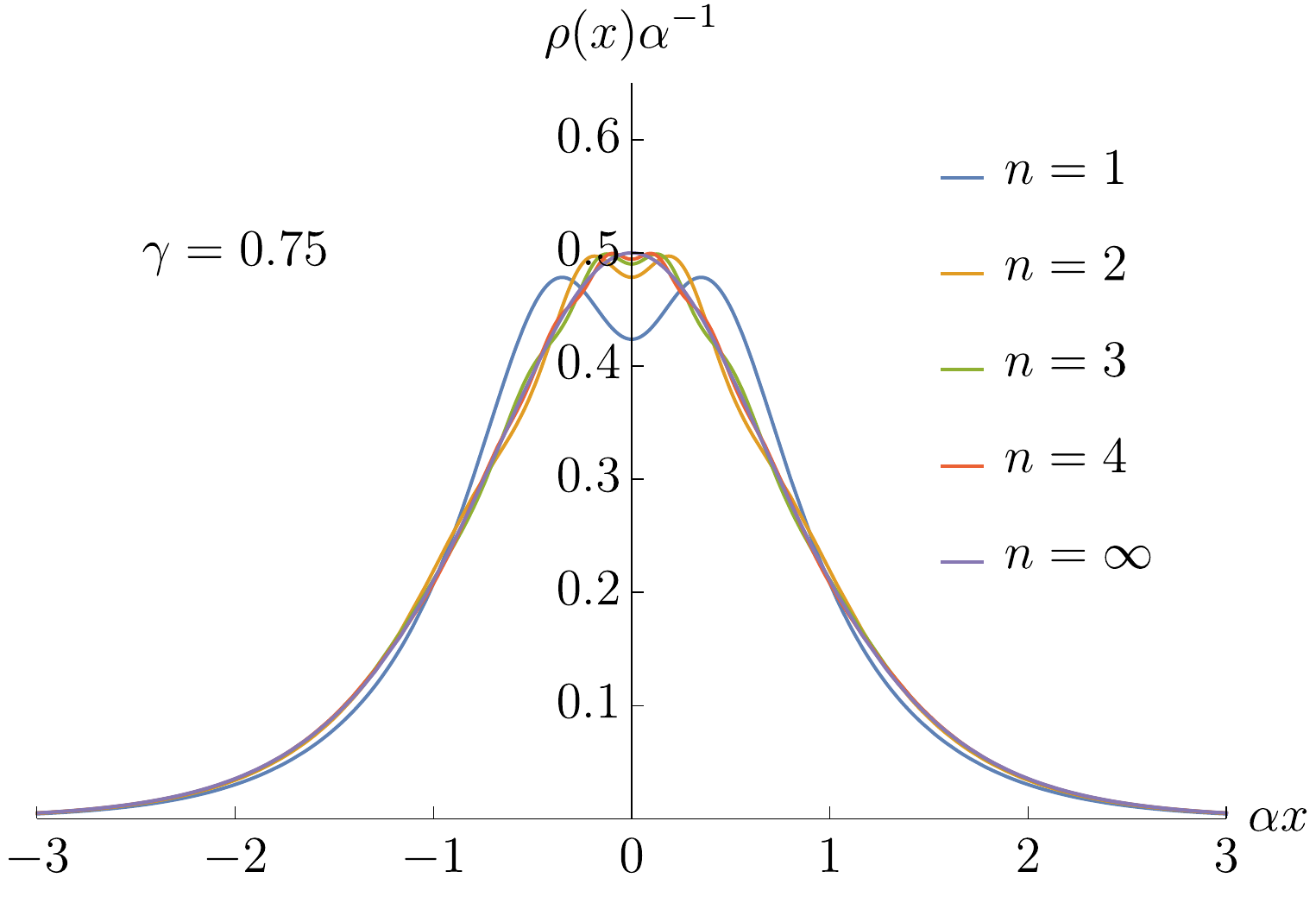}
\includegraphics[scale=0.525]{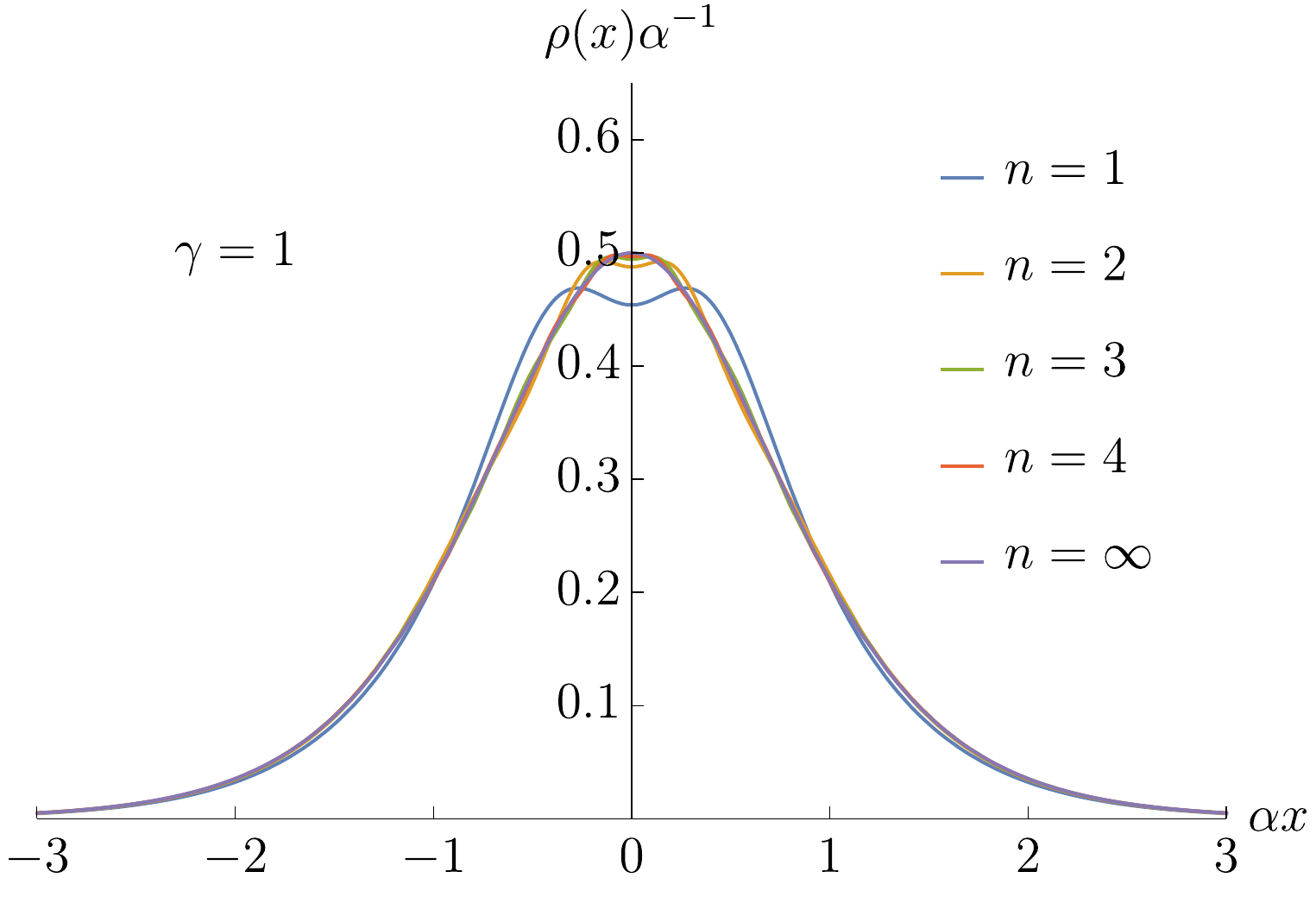}
\caption{\label{Fig:1}Probability density for the discrete levels given in Eq.~\protect\eqref{discretelevels}, $\rho_n(x) \alpha^{-1}$. Shown are the first few excited levels ($n=1,2,3$), as function of the position $x$ (measured in units of $\alpha^{-1}$) for different values of the parameter $\gamma:=0.25,0.5,0.75,1$.  }
\end{figure*}
The normalization constant can be determined from the relation 
\beq
\int _{-\infty}^\infty \psi_n^\dag\psi_n^{\phantom{\dag}} dx=\int _{-\infty}^\infty\left( \psi_{n1}^{*} \psi_{n1}^{\phantom{\dag}}+ \psi_{n2}^{*} \psi_{n2}^{\phantom{\dag}} \right) dx =1
\eeq
and is given by
\beq
{\cal N}_n = \sqrt{ \frac{ \alpha}{2}} \left[ 1+ (n \pi\gamma)^2  \right]^{-\frac{1}{2}}                        
\eeq
From the wave-function given in Eq.~(\ref{discretelevels}) one can easily deduce the probability density $\rho_n(x)= \psi_{1n}^{*} \psi_{1n}^{\phantom{*}}+ \psi_{2n}^{*} \psi_{2n}^{\phantom{*}}$.   
We find:
\[
\rho_n(x) = \frac{\alpha}{2}\,\frac{\text{sech}^2\alpha x}{[1+(n\pi\gamma)^2]}
\left[2\sin^2(n\pi\tanh(\alpha x))+(n\pi\gamma)^2 \right]
\]
and we observe that for large values of the quantum number $n$ the probability density becomes independent of $n$:
\begin{equation}
\lim_{n\to \infty}  \rho_n(x) = \frac{\alpha}{2}\,\,\text{sech}^2\alpha x
\end{equation}
and, depending on the numerical value of $\gamma$, deviations from this limiting form will be appreciable only for the first excited levels. Note that the wave-functions in Eq.~(\ref{discretelevels}) actually develop nodes (zeros of the probability density) only if $\gamma=0$.
Such behaviour is depicted in Fig.~\ref{Fig:1} where we plot the probability density for the first excited states and for different choices of the parameter $\gamma$.

For a stationary state the probability density is time independent and thus the continuity equation dictates that the current satisfies $\partial j /\partial x =0$, or that $j(x,t)$ be a constant in space.  Indeed for the solutions of the discrete levels  given by Eq.~\eqref{discretelevels}
the current  can be easily derived, using the shorthand notation $\Lambda_n(x) = n\pi \tanh(\alpha x)$, as:
\begin{eqnarray}
\label{currentdiscrete}
j_n(x,t)&=& N_n^2 \sqrt{v_0} \left\{ \sin^2[\Lambda_n]-[1+(n\pi\gamma)^2]\sin^2[\Lambda_n]\right.\nonumber\\ && \left.\phantom{xxxxxxxxxxxxx}-(n\pi\gamma)^2\cos^2[\Lambda_n]\right\}\nonumber \\
&=& -\frac{\alpha v_0}{2}\, \frac{(n\pi\gamma)^2}{1+(n\pi\gamma)^2}\,.  
\end{eqnarray}

Let us now examine the situation for $E^2<|\Delta|^2$. In this case the solution of Eq.(\ref{phi1}) is given by
\begin{eqnarray}\label{gr}
\phi_1^<(q)&=&d_1~\text{sinh}\left(\sqrt{\f{|\Delta|^2-E^2}{\hbar^2}}q\right)\phantom{xxxx}\nonumber \\ && \phantom{xxxxxxxx}+d_2~\text{cosh}\left(\sqrt{\f{|\Delta|^2-E^2}{\hbar^2}}q\right)
\end{eqnarray}
where $d_{1,2}$ are arbitrary constants. Clearly there are no acceptable solutions satisfying the boundary conditions in Eq.~(\ref{bc}) and so we conclude that  discrete bound states exist only in the region $E^2>|\Delta|^2$. 

We now discuss bound states in continuum (BIC) solutions. 
 It may be recalled that unlike discrete bound states BIC are states which are normalizable for any value of the energy~\cite{Neumann:1929aa,Stillinger:1975aa}. Here also we consider the region $E^2>|\Delta|^2$ and to obtain BIC within the present framework we consider a formal solution of Eq.(\ref{phi1}), namely
\beq
\phi_{1}^>(q)=N_{>}~\sin(\eps q)
\eeq
where $N_{>}$ is a constant to be determined later by normalisation of the full spinor $\psi$ and $\eps$ can assume any value. Now using Eqs.~(\ref{psiphi}\&\ref{intert2}) and  upon defining:  \[\mu= \sqrt{\frac{E^2}{|\Delta|^2 }-1} \qquad \qquad \mu \in [0, +\infty]\] we find:
\begin{widetext}
\beq
\label{BIC1}
\psi^{>}=\left(
\ba{l}
\psi_1^{>}(x)\\ \psi_2^{>}(x) \end{array}\right)= {\cal N}_{>}\,\,\text{sech}  (\alpha x)\,\, \left(\begin{array}{c}\sin \left[\frac{\mu}{\gamma}\, \tanh(\alpha x)\right]  \\ 
  \frac{E}{\Delta} \sin \left[\frac{\mu}{\gamma}\, \tanh(\alpha x) \right]+i \frac{|\Delta|}{\Delta} \mu \cos \left[\frac{\mu}{\gamma}\, \tanh(\alpha x) \right] \\
\ea
\right)
\eeq
\begin{figure*}[t]
\includegraphics[scale=0.525]{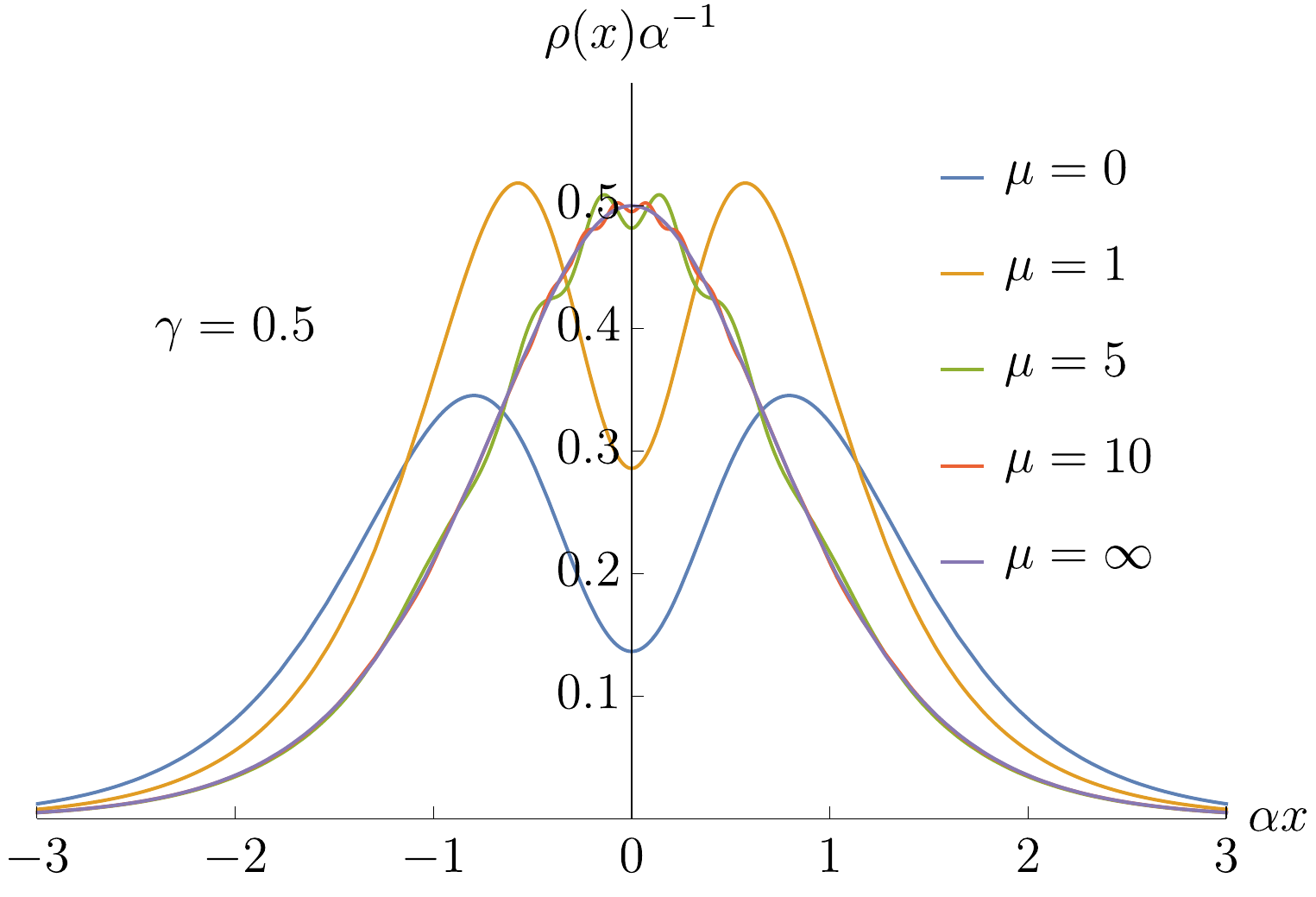}
\includegraphics[scale=0.525]{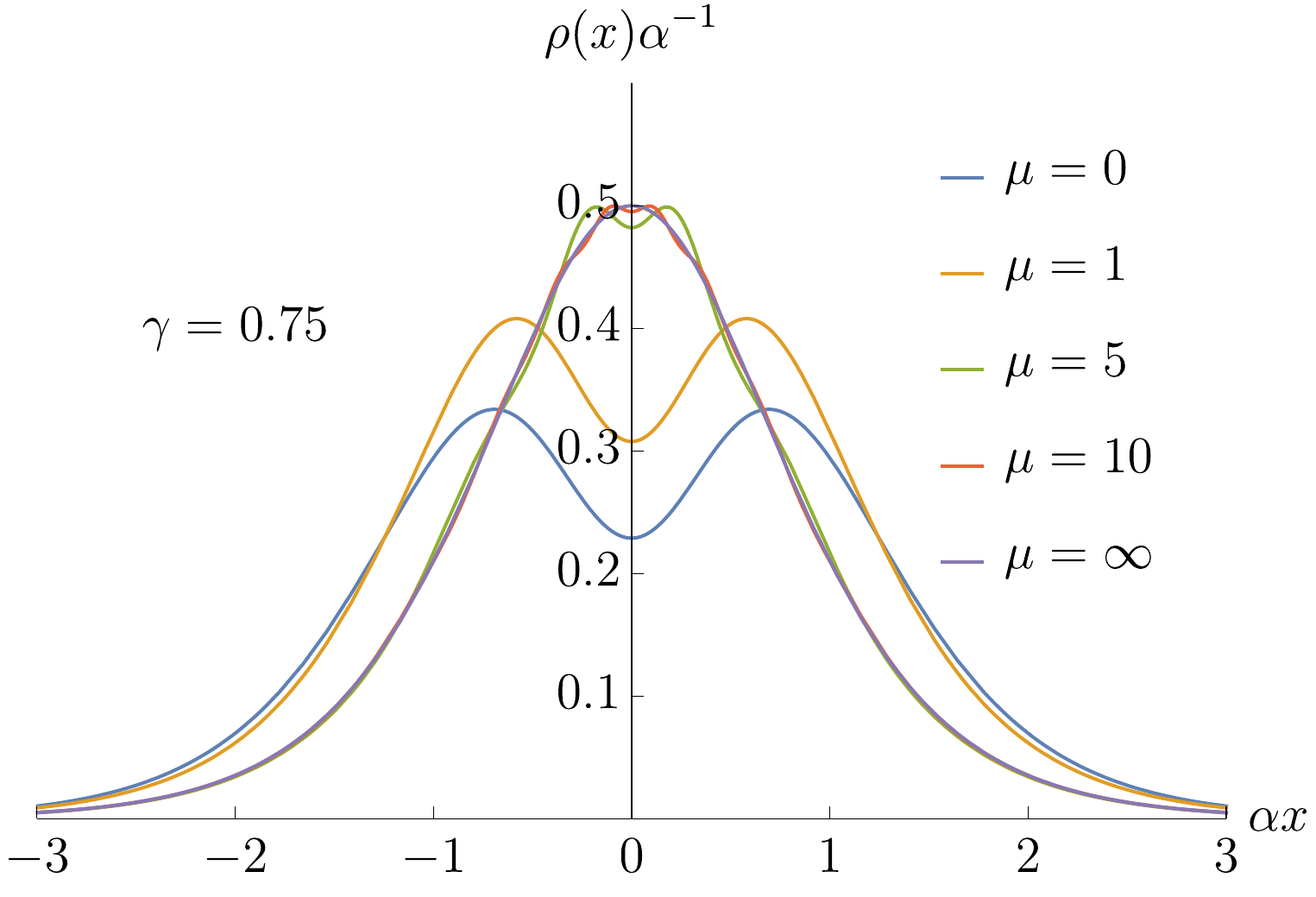}
\caption{\label{Fig:2}Probability density ($\rho_>(x) \alpha^{-1}$) for the BIC solutions in the region $E>|\Delta|$ given in Eq.~\protect\eqref{BIC1}. Shown are the probability densities for various values of the dimensionless parameter $\mu$, as function of the position $x$ (measured in units of $\alpha^{-1}$) for different values of the parameter $\gamma:=0.5$ (left) and $\gamma:=0.75$ (right). }
\end{figure*}
\end{widetext}
From the above expressions it may be observed that since $sine$ and the $cosine$ can not exceed unity the sech($\alpha x$) term in the wave functions ensures that they decay to $0$ as $x\ra\pm\infty$. 

In fact the normalization constant can be exactly evaluated and is given by
\beq
{\cal N}_{>}^{-1}=\sqrt{\frac{2}{\alpha}}\, \sqrt{(1+\mu^2) -\frac{\gamma}{2\mu}\, \sin\left(\frac{2\mu}{\gamma}\right)}
\eeq
The probability density is found to be given as:
\begin{equation}
\label{rho0p}
\rho_>(x)= \frac{\alpha}{2}\,\,\frac{\text{sech}^2(\alpha x)
\left[\mu^2 +2 \sin^2\left(\frac{\mu}{\gamma} \tanh(\alpha x)\right)\right]
}{(1+\mu^2) -\frac{\gamma}{2\mu}\, \sin\left(\frac{2\mu}{\gamma}\right)}\,\,
\end{equation}
We  note  that for   values of the energy close to $|\Delta|$ or  $\mu \to 0$,  the probability density approaches a limiting shape that depends only on the parameter $\gamma$:
\begin{equation}
\label{rho0plim}
\lim_{\mu\to 0^+}  \rho_>(x) = \frac{\alpha\, \text{sech}^2(\alpha x)}{2(1+\frac{2}{3\gamma^2})}\,\,\left(1+\frac{2}{\gamma^2}\tanh^2\alpha x \right).
\end{equation} Note that the explicit limiting form of $\rho_>(x)$ given in Eq.~\eqref{rho0plim} is easily checked to be exactly normalised to unity. 

The current $j_>$ is computed using the wave functions in Eq.~\eqref{BIC1} into the general expression given in Eq.~\eqref{current}.  A straightforward computation gives:
\begin{equation}
\label{currentp}
j_>= - \frac{\alpha v_0}{2} \,\, \frac{\mu^2}{1+\mu^2 -\frac{\gamma}{2\mu}\sin\left(\frac{2\mu}{\gamma}\right)}\,.
\end{equation} 
In Fig.~\ref{Fig:2}  we  plot the probability density $\rho_>(x)$ for different sets of parameter values thus confirming our finding.
\begin{figure*}[t]
\includegraphics[scale=0.525]{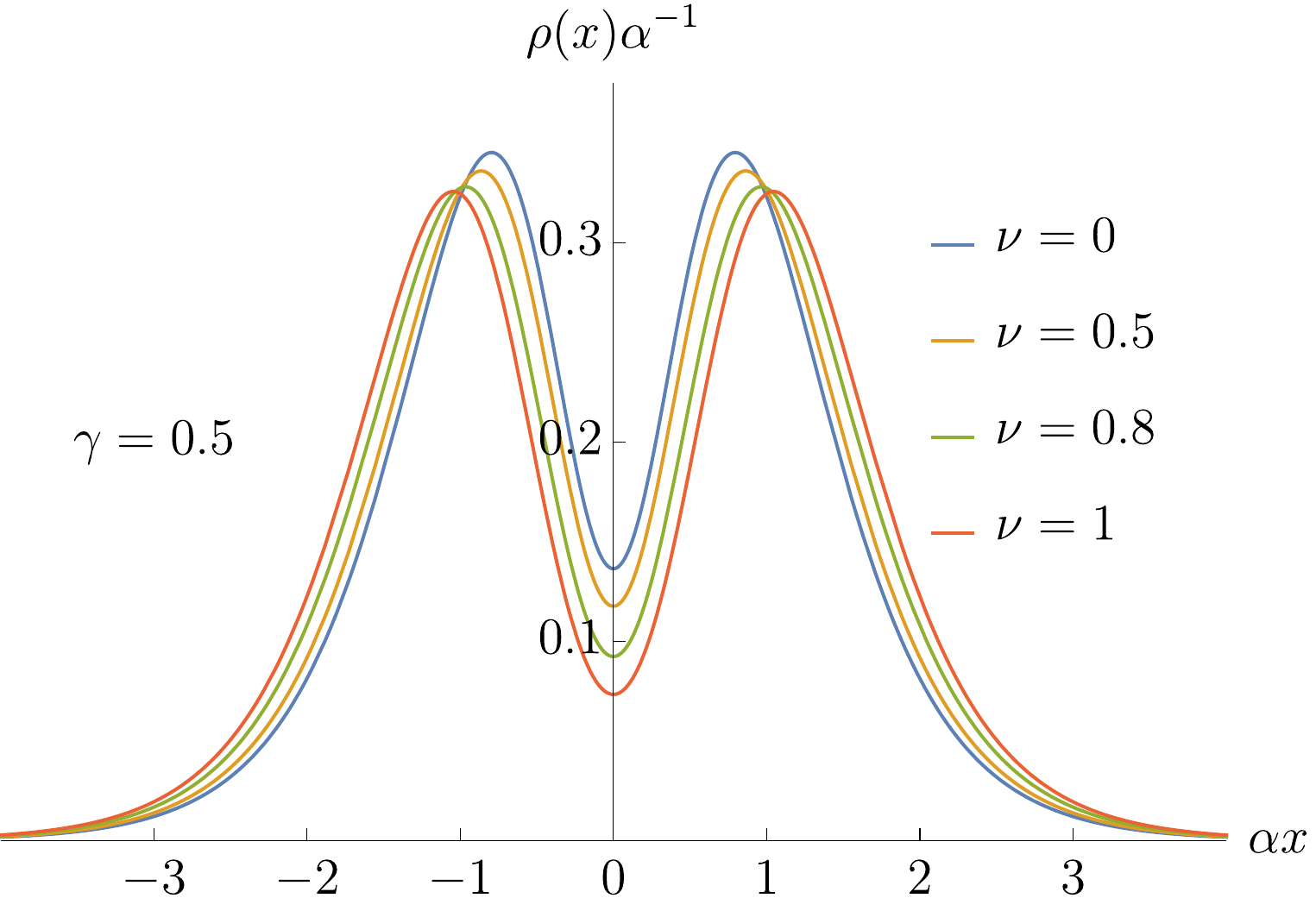}
\includegraphics[scale=0.525]{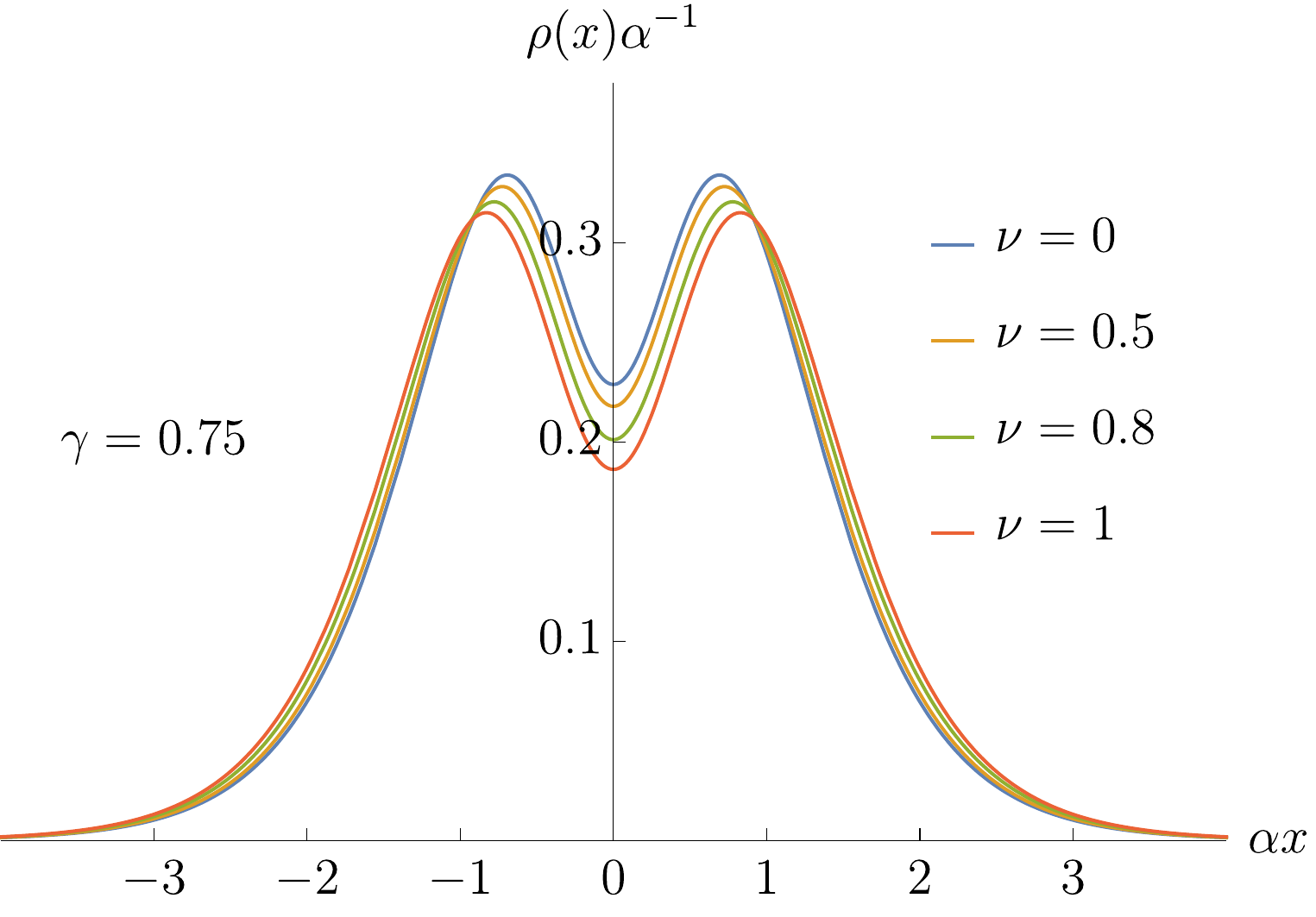}
\caption{\label{Fig:3}Probability density ($\rho_<(x) \alpha^{-1}$) for the BIC solutions in the region $E>|\Delta|$ given in Eq.~\protect\eqref{BIC2}. Shown are the probability densities for various values of the dimensionless parameter $\nu$, as function of the position $x$ (measured in units of $\alpha^{-1}$) for different values of the parameter $\gamma:=0.5$ (left) and $\gamma:=0.75$ (right). }
\end{figure*}
Next we consider the nature of the solution in the region $E^2<|\Delta|^2$. In this case the solution is again given by Eq.~(\ref{gr}). For the sake of simplicity we put $d_2=0$ and the solution then becomes
\beq
\phi_1^{<}(x)=d_1~\text{sinh}\left(\sqrt{\f{|\Delta|^2-E^2}{\hbar^2}}\f{\text{tanh}(\a x)}{\a v_0}\right) 
\eeq
Then upon defining the dimensionless parameter:
\begin{equation}
\nu= \sqrt{1-\frac{E^2}{|\Delta|^2}},\qquad\qquad \nu \in [0,1]
\end{equation} using Eqs.~(\ref{psiphi}\&\ref{intert2}) we finally obtain:
\begin{widetext}
\beq
\label{BIC2}
\psi^{<}=\left(
\ba{l}
\psi_1^{<}(x)\\ \psi_2^{<}(x) \end{array}\right)= {\cal N}_{<}\,\,\text{sech}  (\alpha x)\,\, \left(\begin{array}{c}\text{sinh} \left[\frac{\nu}{\gamma}\, \tanh(\alpha x)\right]  \\ 
  \frac{E}{\Delta} \text{sinh} \left[\frac{\nu}{\gamma}\, \tanh(\alpha x) \right]+i \frac{|\Delta|}{\Delta} \nu \cosh \left[\frac{\nu}{\gamma}\, \tanh(\alpha x) \right] \\
\ea
\right)\, .
\eeq
\end{widetext}

It is not difficult to see that both of $\psi_{1,2}^<$ go to zero as $x\ra\pm\infty$. In fact the normalization constant can be found to be
\beq
{\cal N}_{<}^{-1} = \sqrt{\frac{2}{\alpha}}\sqrt{\frac{\gamma}{2\nu}\text{sinh}\left(\frac{2\nu}{\gamma}\right)-(1-\nu^2)}\eeq

\begin{figure}[h]
\includegraphics[scale=0.55]{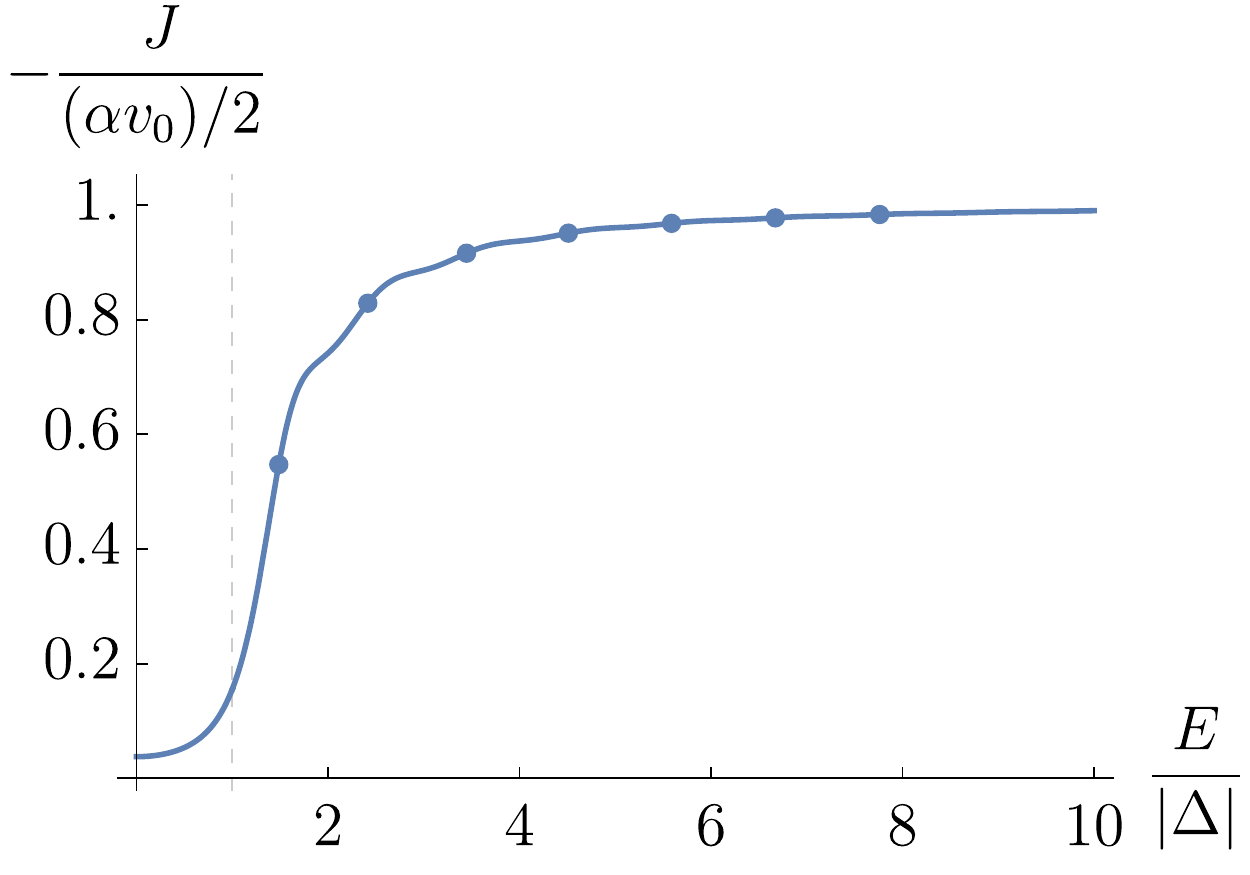}
\caption{\label{Fig:4}Minus the probability current in units of $\alpha v_0/2$, $-j/(\alpha v_0 /2)$, for both discrete bound states and BIC solutions as function of the energy $E$ in units of the gap parameter $|\Delta|$. The BIC current extends to the region $E<|\Delta|$ while in the region  $E\ge|\Delta|$ we  show both the currents of the BIC-like solution (continuos line) as well as the currents of the discrete bound states (full dots).}
\end{figure}

From the wave-function given in Eq.~(\ref{BIC2}) one can easily deduce the probability density $\rho_<(x)= \psi_{1}^{<*} \psi_{1}^{<\phantom{*}}+ \psi_{2}^{<*} \psi_{2}^{<\phantom{*}}$.   
We find:
\begin{equation}
\label{rho0m}
\rho_<(x) = \frac{\alpha}{2}\,\frac{\text{sech}^2\alpha x \left\{2\sinh^2\left[\frac{\nu}{\gamma}\tanh(\alpha x)\right]+\nu^2 \right\}}
{\frac{\gamma}{2\nu}\sinh\left(\frac{2\nu}{\gamma}\right)-(1-\nu^2)}
\,.
\end{equation}
The current $j_<$ is again computed using the wave functions in Eq.~\eqref{BIC2} into the general expression given in Eq.~\eqref{current}.  A straightforward computation now gives:
\begin{equation}
\label{currentm}
j_<= - \frac{\alpha v_0}{2} \,\, \frac{\nu^2}{\frac{\gamma}{2\nu}\text{sinh}\left(\frac{2\nu}{\gamma}\right)-(1-\nu^2)}\,.
\end{equation} 
In Fig.~\ref{Fig:3} we have plotted the probability density $\rho_<(x)$ which clearly exhibits the BIC nature of the wave functions (note that the parameter $\nu$ can be changed countinously in the interval $[0,1]$ ). We  observe that for vanishing  values of the energy, $E\to 0$ or $\nu \to 0$,  the probability density approaches a shape that depends only on the parameter $\gamma$ and is therefore independent of the energy:
\begin{eqnarray}
\label{rho0mlimit}
\lim_{\nu\to 0}  \rho_<(x) &=& \lim_{\mu\to 0^+}  \rho_>(x) \nonumber \\ &=&\frac{\alpha\,\text{sech}^2(\alpha x)}{2(1+\frac{2}{3\gamma^2})}\,\,\left(1+\frac{2}{\gamma^2}\tanh^2\alpha x \right).
\end{eqnarray} Note that the explicit limiting form of $\rho_<(x)$ given in Eq.~\eqref{rho0mlimit} is easily seen to coincide with the limiting form obtained in Eq.~\eqref{rho0plim} and thus again to be exactly normalised to unity. Thus the BdG equation possesses BIC solutions in both the normal phase,  $E>|\Delta|$ and the superconducting phase, $E\le|\Delta|$. In Fig.~\ref{Fig:4} we show the probability currents as function of the energy in units of the gap parameter $|\Delta|$ ($E/|\Delta|$) where we combine the results found above about the probability currents, c.f., Eqs.~(\ref{currentdiscrete},\ref{currentp},\ref{currentm}). \vspace{0.2cm}\\
We remark that with the specific choice in Eq.~\ref{v1} of the velocity profile the solutions described above, discrete bound states and bound states in continuum (BIC) exhaust all solutions. There are no scattering states. This is not surprising since the velocity profile in Eq.~\ref{v1} has been chosen so, just to be able to find bound states and BIC solutions. In order to have scattering solutions different velocity and/or gap profiles must be chosen. We will take up this point in the next paragraph.

\vspace{0.15cm}
\noindent\underline{Step-like Fermi velocity and gap profiles}\\
It is well known that it is not entirely trivial to describe non zero reflectance within the  massless Dirac equation. For instance  ref.~\cite{Peres:2009aa} discusses a massless Dirac Hamiltonian with a position dependent Fermi velocity $v_F(x)$ finding that a simple step-like profile is not enough to get non zero reflectance. Here our aim is to show that with the Hamiltonian given in Eq.~\eqref{HvFx} introducing a step-like gap parameter allows scattering states with \emph{non vanishing  reflectance}. We  take up the following profiles: 
\begin{subequations}
\begin{align}
v_F(x) &= v_F^- \,\theta(-x) +v_F^+\, \theta(x)\, , \\
\Delta(x) & = \Delta^{\!\!-}\, \theta(-x) +\Delta^{\!\!+}\, \theta(x)\, .
\end{align}
\end{subequations}  
In particular  we show that it is enough to assume a non vanishing phase difference  between $\Delta^+$ and $\Delta^-$ to trigger normal reflectance. 

Let us consider first the case $E^2>\text{max}\left\{\lvert \Delta^{\!\!-}\rvert^2,\lvert \Delta^{\!\!+}\rvert^2\right\}$. 
We note that integrating the first order intertwining equations in Eq.~\eqref{intert2} over an infinitesimal interval around $x=0$ one finds the following boundary conditions:
\begin{subequations}
\label{bc2}
\begin{align}
\label{bc2a}\phi_1(0^-)&=\phi_1(0^+)\\
\label{bc2b}\phi_2(0^-)&=\phi_2(0^+)
\end{align}
\end{subequations}  
Then one can easily solve the corresponding second order equations in Eq.~\eqref{secondordereq}. Upon defining: 
\begin{equation}
k_{-}^2=\frac{\sqrt{E^2- \lvert \Delta^{\!\!-} \rvert^2}}{(\hbar v_F^-)^2}\qquad
k_{+}^2=\frac{\sqrt{E^2- \lvert \Delta^{\!\!+} \rvert^2}}{(\hbar v_F^+)^2}
\end{equation}
the following scattering solution is found: 
\[
\phi_1(x)  =
\left\{
\!
\begin{aligned}
A e^{ik_-x} +B e^{-ik_-x}  &\,\, \text{ if }\,\, x<0\\
C e^{ik_+x} & \,\, \text{ if }\,\, x>0
\end{aligned}
\right.
\]

The lower component $\phi_2(x)$ is obtained using the intertwining relations in Eq.~\eqref{intert2}:
\begin{widetext}
\[
\phi_2(x)  =
\left\{
\!
\begin{aligned}
A\left(\frac{E}{\Delta^{\!\!-}}-\frac{\hbar v_F^-k_-}{\Delta^{\!\!-}}\right) e^{+ik_-x} +B\left(\frac{E}{\Delta^{\!\!-}}+\frac{\hbar v_F^-k_-}{\Delta^{\!\!-}}\right) e^{-ik_-x}  &\,\, \text{ if }\,\, x<0\\
C\left(\frac{E}{\Delta^{\!\!+}}-\frac{\hbar v_F^+k_+}{\Delta^{\!\!+}}\right) e^{+ik_+x} & \,\, \text{ if }\,\, x>0
\end{aligned}
\right.
\]
\end{widetext}
 We implement the boundary conditions in Eqs.~(\ref{bc2a}\&\ref{bc2b}) in the above solutions $\phi_{1,2}(x)$ and find: 
\begin{subequations}
\label{BA}
\begin{align}\label{BAa}
B&=A \frac{\displaystyle\frac{E}{\Delta^{\!\!-}}-\frac{E}{\Delta^{\!\!+}}-\frac{\hbar v_F^-k_-}{\Delta^{\!\!-}}+\frac{\hbar v_F^+k_+}{\Delta^{\!\!+}}}{\displaystyle\frac{E}{\Delta^{\!\!+}}-\frac{E}{\Delta^{\!\!-}}-\frac{\hbar v_F^-k_-}{\Delta^{\!\!-}}-\frac{\hbar v_F^+k_+}{\Delta^{\!\!+}}}\\ \label{BAb} C&=A+B
\end{align}
\end{subequations}
We note that if $\Delta(x)=\text{costant}$ (i.e. $\Delta^{\!\!-}=\Delta^{\!\!+}$) there is no reflection because  the coefficient $B$ vanishes identically even if $v_F^- \neq v_F^+$. 

\begin{figure}[t]
\includegraphics[scale=0.525]{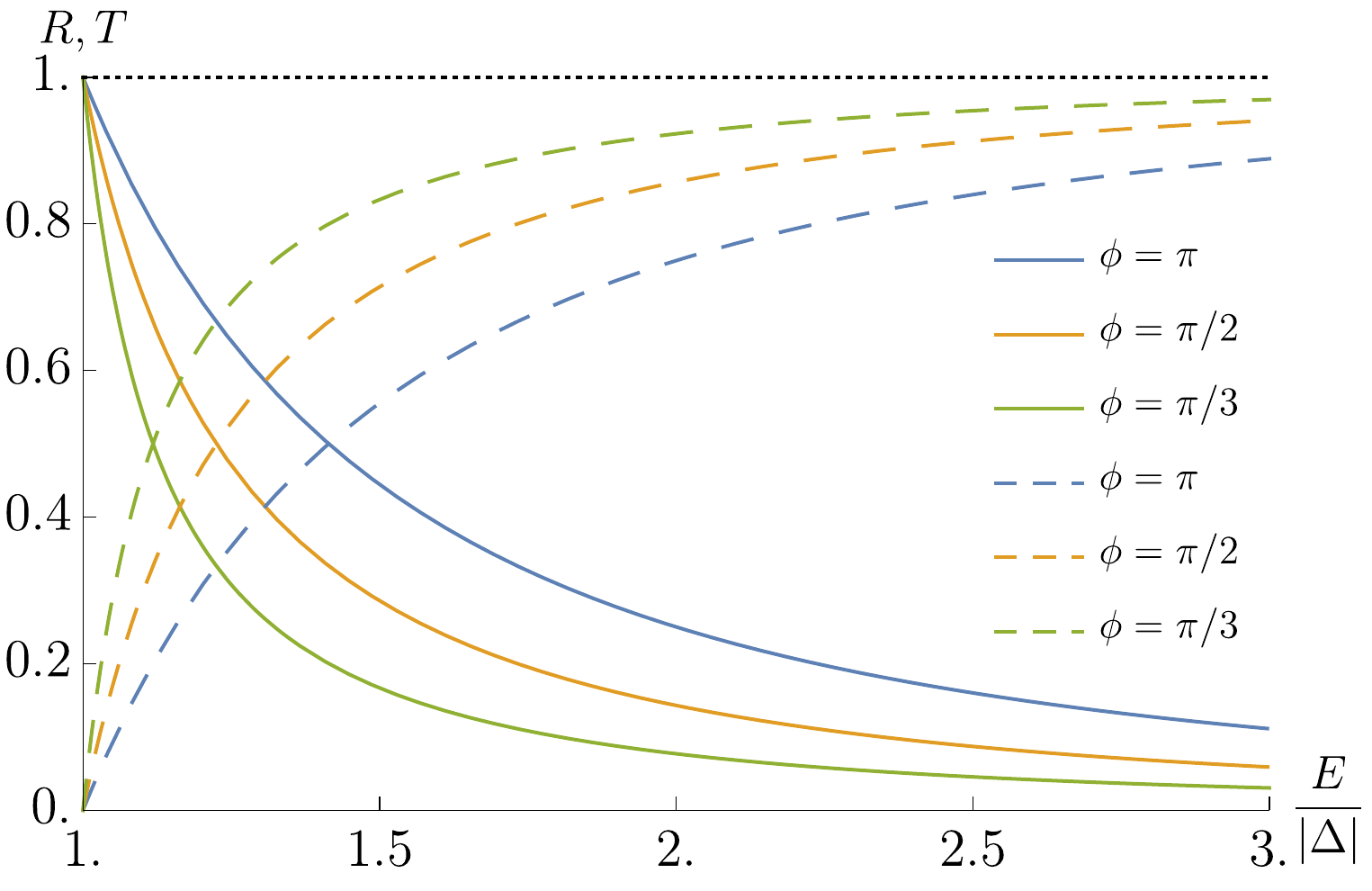}
\caption{\label{Fig:5} Reflectance ($R$) and transmittance ($T$) coefficients for the step-like gap profile with a phase difference between $\Delta^{\!\!-}$ and  $\Delta^{\!\!+}$:  $\Delta^{\!\!+} =\Delta$ and $\Delta^{\!\!-} =\Delta e^{i\phi}$. The solid lines depict the reflectance $R$ for $\phi=\pi$ (blue line), $\phi=\pi/2$ (orange line), $\phi=\pi/3$ (green line). The dashed lines are the transmittance curves with the same color codes. 
We see that unitarity, $R+T=1$, is verified (black dotted line).}
\end{figure}

The probability current is then easily found from $j(x)=\lvert \phi_1\rvert^2 - \lvert \phi_2\rvert^2$ as:
\[
j(x)  =
\left\{
\!
\begin{aligned}
j_{\text{inc.}}-j_{\text{refl.}} &\,\, \text{ if }\,\, x<0\\
j_{\text{trans.}} & \,\, \text{ if }\,\, x>0
\end{aligned}
\right.
\]
with:
\begin{subequations}
\label{fluxes}
\begin{align}
\label{fluxesa} j_{\text{inc.}}&=\lvert A \rvert^2 \left[ 1- \frac{(E-\hbar v_F^-k_-)^2}{\lvert\Delta^{\!\!-}\rvert^2}\right]\\
\label{fluxesb}j_{\text{refl.}}&=\lvert B \rvert^2 \left[  \frac{(E+\hbar v_F^-k_-)^2}{\lvert\Delta^{\!\!-}\rvert^2}-1\right]\\
\label{fluxesc} j_{\text{refl.}}&= \lvert C \rvert^2 \left[ 1- \frac{(E-\hbar v_F^+k_+)^2}{\lvert\Delta^{\!\!+}\rvert^2}\right]
\end{align}
\end{subequations}
So that the reflectance ($R$) and transmittance ($T$) coefficients can be deduced as:
\begin{subequations}
\label{RT}
\begin{align}
\label{RTa}R &=\frac{j_{\text{refl.}}}{j_{\text{inc.}}}= \frac{\lvert B\rvert^2}{\lvert A\rvert^2}\, \frac{-\lvert\Delta^{\!\!-}\rvert^2+(E+\hbar v_F^- k_-)^2}{+\lvert\Delta^{\!\!-}\rvert^2-(E-\hbar v_F^- k_-)^2}\\
\label{RTb}T &= \frac{j_{\text{trans.}}}{j_{\text{inc.}}}= \frac{\lvert C\rvert^2}{\lvert A\rvert^2}\, \frac{\lvert\Delta^{\!\!-}\rvert^2}{\lvert\Delta^{\!\!+}\rvert^2}\, \frac{\lvert\Delta^{\!\!+}\rvert^2-(E-\hbar v_F^+ k_+)^2}{\lvert\Delta^{\!\!-}\rvert^2-(E-\hbar v_F^- k_-)^2}
\end{align}
\end{subequations} 
With the help of Eqs.~(\ref{BAa}\&\ref{BAb}) the reflectance $R$ and transmittance $T$ can be explicitly computed as functions of the energy eigenvalue $E$. In Fig.~\ref{Fig:5} we show $R$ and $T$ for a particular model of steplike gap: $\Delta^{\!\!+} =\Delta$ and $\Delta^{\!\!-} =\Delta e^{i\phi}$. We provide curves for $R$ and $T$ as functions of $E/\lvert\Delta\rvert$ for different values of the phase difference ($\phi=\pi, \pi/2,\pi/3$).

\begin{figure}[h]
\includegraphics[scale=0.525]{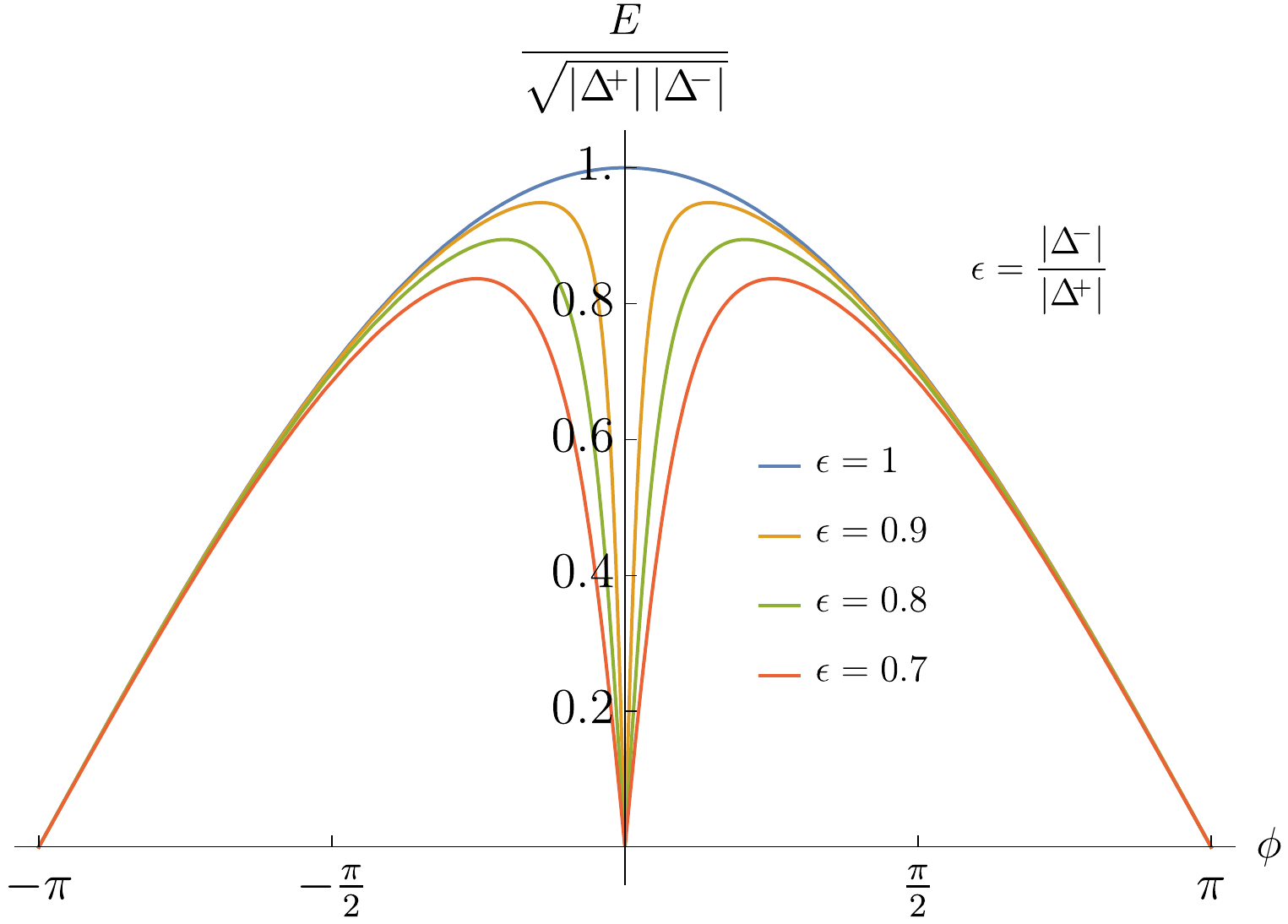}
\caption{\label{Fig:6} Positive branch of the energy eigenvalue obtained merging the  two solutions of Eq.~\eqref{eqeigenvalue} given in Eq.~\eqref{generalBS} for different values of the parameter $\epsilon=\lvert \Delta^{\!\!+}\rvert / \lvert \Delta^{\!\!-}\rvert$. Note that the solutions in Eq.~\eqref{generalBS} are  symmetric under the exchange $\lvert \Delta^{\!\!+}\rvert \leftrightarrow \lvert \Delta^{\!\!-}\rvert$ (or $\epsilon \leftrightarrow 1/\epsilon$).}
\end{figure}

We now discuss the region $E^2 < \text{min}\left\{\lvert \Delta^{\!\!-}\rvert^2,\lvert \Delta^{\!\!+}\rvert^2\right\}$
and we show that it admits a single bound state. Here we can define:
\begin{equation}
\label{kappas}
\kappa_{-}^2=\frac{\sqrt{\lvert \Delta^{\!\!-} \rvert^2-E^2}}{(\hbar v_F^-)^2}\qquad
\kappa_{+}^2=\frac{\sqrt{\lvert \Delta^{\!\!+} \rvert^2-E^2}}{(\hbar v_F^+)^2}
\end{equation}
We find that Eq.~\eqref{secondordereq} admits the solution:
\[
\phi_1(x)  =
\left\{
\!
\begin{aligned}
B\, e^{-i\kappa_-x}  &\,\, \text{ if }\,\, x<0\\
C\, e^{+i\kappa_+x} & \,\, \text{ if }\,\, x>0
\end{aligned}
\right.
\]
and the corresponding lower component $\phi_2(x)$ is found via Eq.~\eqref{intert2a}:
\[
\phi_2(x)  =
\left\{
\!
\begin{aligned}
B\,\frac{E+i\hbar v_F^-\kappa_-}{\Delta^{\!\!-}}\, e^{-i\kappa_-x}  &\,\, \text{ if }\,\, x<0\, ,\\
C\,\frac{E-i\hbar v_F^+\kappa_+}{\Delta^{\!\!+}}\, e^{+i\kappa_+x} & \,\, \text{ if }\,\, x>0\, .
\end{aligned}
\right.
\]
The boundary conditions in Eqs.~(\ref{bc2a}\&\ref{bc2b}) are easily found to admit a non vanishing solution for $B$ and $C$ if and only if:
\begin{equation}
\label{eqeigenvalue}
\frac{E+i\hbar v_F^-\kappa_-}{\Delta^{\!\!-}} = \frac{E-i\hbar v_F^+\kappa_+}{\Delta^{\!\!+}}
\end{equation}
which is a second order equation, see Eq.~\eqref{kappas}, for the eigenvalue $E$ of  the bound state. The two solutions are:
\begin{equation}
\label{generalBS}
E= \frac{\pm{\sqrt{\lvert \Delta^{\!\!+}\rvert\,\, \lvert \Delta^{\!\!-}\rvert}}\,\sin\phi\, \sqrt{2\cos\phi-\frac{\lvert \Delta^{\!\!-}\rvert}{\lvert \Delta^{\!\!+}\rvert}-\frac{\lvert \Delta^{\!\!+}\rvert}{\lvert \Delta^{\!\!-}\rvert}}}{\sqrt{\left[ 2 \cos\phi \frac{\lvert \Delta^{\!\!-}\rvert}{\lvert \Delta^{\!\!+}\rvert} -1 -\frac{\lvert \Delta^{\!\!-}\rvert^2}{\lvert \Delta^{\!\!+}\rvert^2} \right] \left[ 1 +\frac{\lvert\Delta^{\!\!+}\rvert^2}{\lvert \Delta^{\!\!-}\rvert^2}-2\cos\phi\frac{\lvert\Delta^{\!\!+}\rvert}{\lvert \Delta^{\!\!-}\rvert}\right]}}\, .
\end{equation}
The two solutions ($\pm$) are respectively positive definite for $\phi \in \left({ [0,+\pi] \atop [-\pi, 0]} \right)$ and negative definite for $\phi \in \left({[-\pi, 0] \atop  [0,+\pi]}\right)$. Fig.~\ref{Fig:6} shows the merging of the two solutions of Eq.~\ref{generalBS} in the positive branch of the spectrum.  

\begin{figure*}
\includegraphics[scale=0.5]{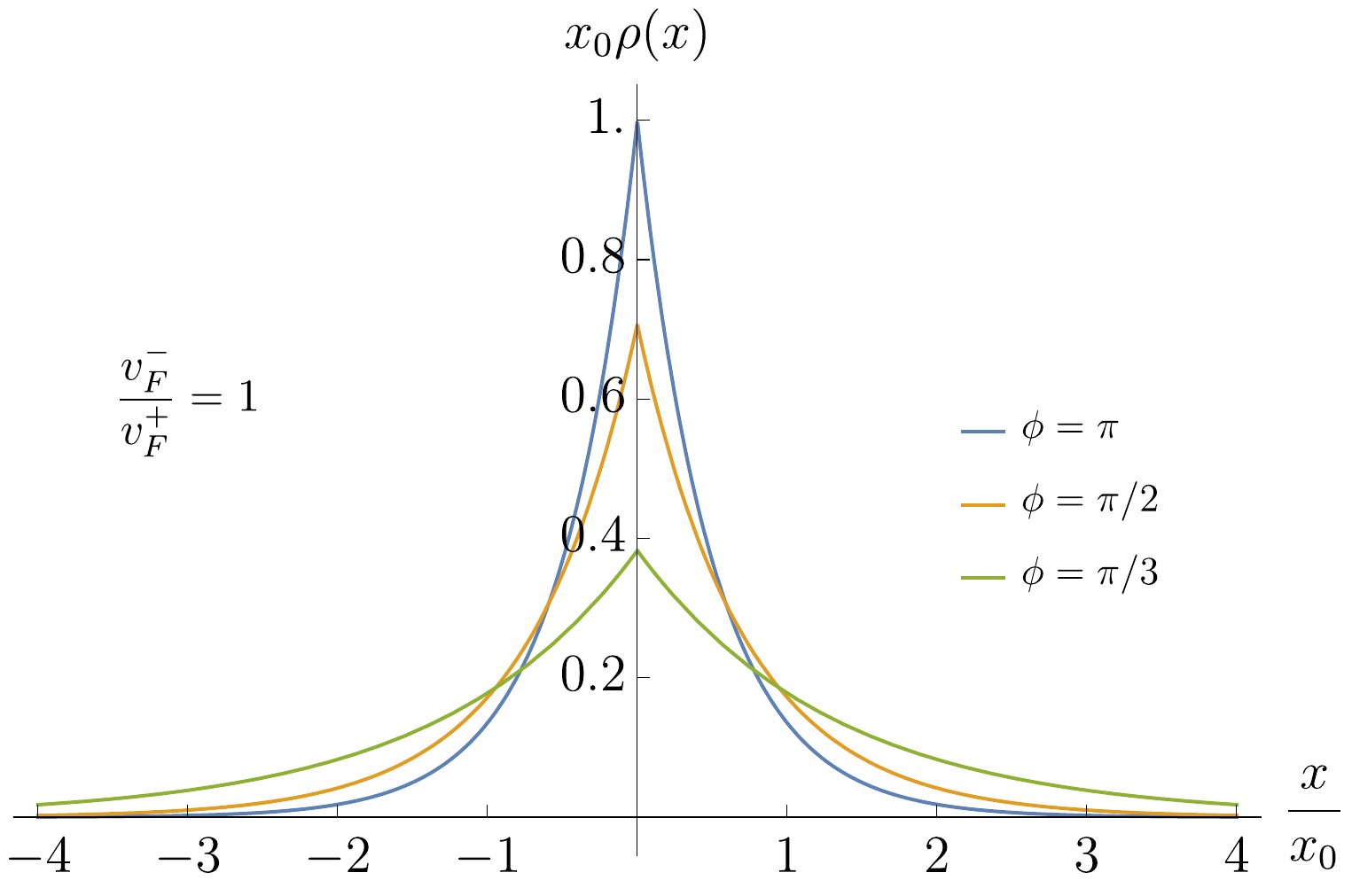}\hspace{1.0cm}
\includegraphics[scale=0.5]{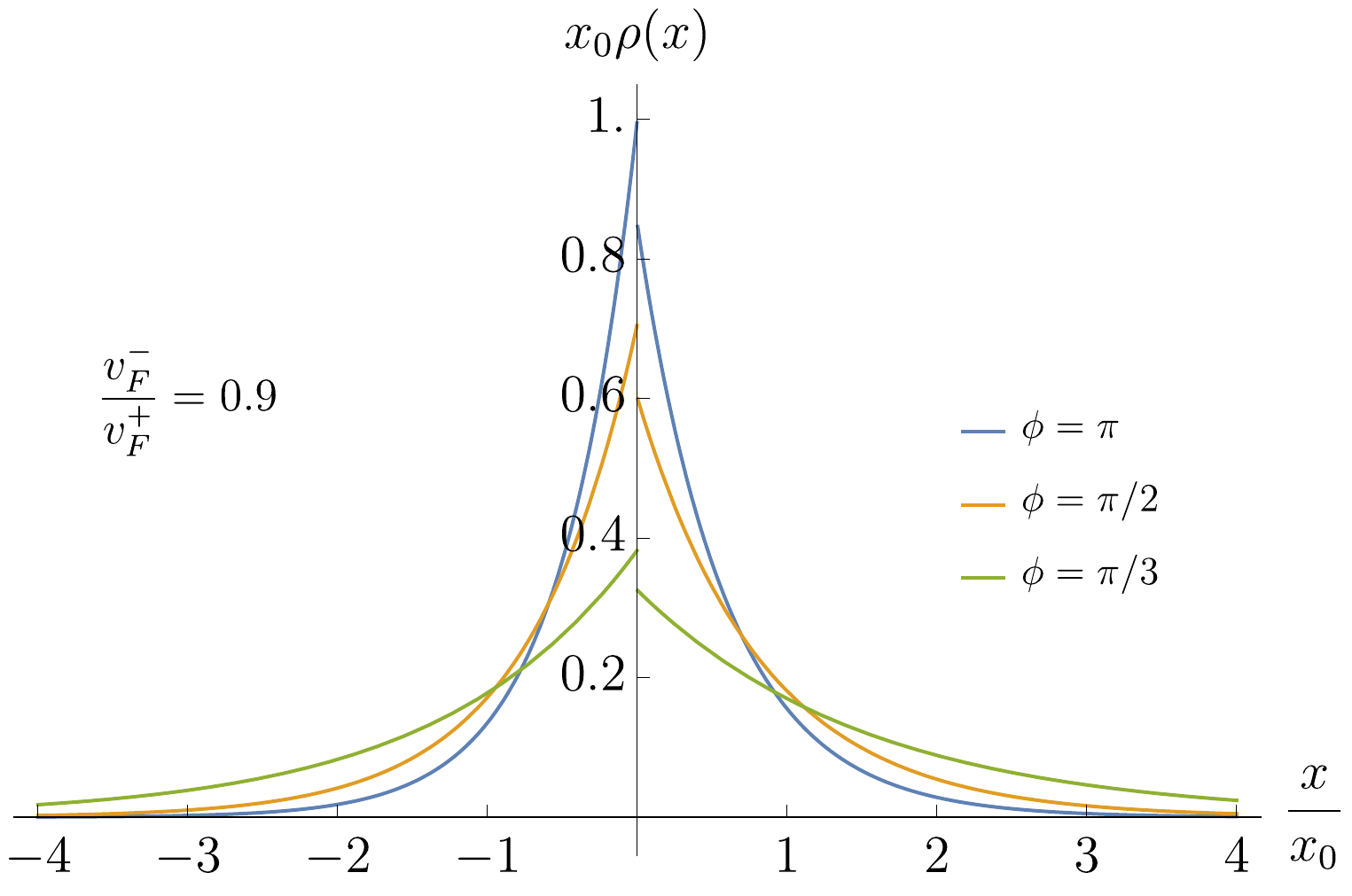}
\caption{\label{Fig:7} Probability density $x_0 \rho(x)$ as a function of $x/x_0$  for the step-like gap profile with a phase difference between $\Delta^{\!\!-}$ and  $\Delta^{\!\!+}$:  $\Delta^{\!\!+} =\Delta$ and $\Delta^{\!\!-} =\Delta e^{i\phi}$.
On the left we show the case $v_F^-=v_F^+$ (the probability density is continuous at $x=0$), while on the plot on the right we have $v_F^-/v_F^+=0.9$ (the probability density shows a discontinuity at $x=0$). In each plot the three curves are for three different values of the phase parameter $\phi$: $\phi=\pi$ (blue line), $\phi=\pi/2$ (orange line), and $\phi=\pi/3$ (green line). }
\end{figure*}

We can then consider the  simpler model  $\Delta^{\!\!+} =\Delta$ and $\Delta^{\!\!-} =\Delta e^{i\phi}$ and find that Eq.~\eqref{generalBS} reduces to:
\begin{equation}
\label{specificBS}
E=\pm \cos\frac{\phi}{2}\, \Delta
\end{equation}
and in the positive branch of the spectrum we have then the single bound state at $E=\cos(\phi/2)\Delta$ (blue line in Fig.~\ref{Fig:6}).
The corresponding spinor wave function $\psi$ can be computed with the help of Eq.~\eqref{psiphi}:
\begin{equation}
\label{spinorwf}
\psi(x)  =
\left\{
\!
\begin{aligned}
\frac{B}{\sqrt{v_F^-}}\,\left (1 \atop e^{-i\phi/2}\right )\, e^{+\kappa_- x} &\,\, \text{ if }\,\, x<0\, ,\\
\frac{B}{\sqrt{v_F^+}}\,\left (1 \atop e^{+i\phi/2}\right )\, e^{-\kappa_+ x} & \,\, \text{ if }\,\, x>0\, .
\end{aligned}
\right.
\end{equation}
We note that given the continuity of the $\phi_{1,2}$ components the corresponding $\psi_{1,2}$ spinor components are in general not continuous (if $v_F^- \ne v_F^+$). This may reflect in a discontinuity in the probability density. The spinor in Eq.~\eqref{spinorwf} is straightforwardly normalised to unity and the corresponding probability density is computed as:
\begin{equation}
\label{probability}
x_0\, \rho(x)  = 
\sin\frac{\phi}{2}
\left\{
\!
\begin{aligned}
 e^{+2\sin(\phi/2) \frac{x}{x_0}} &\,\, \text{ if }\,\, x<0\\
 \frac{v_F^-}{v_F^+}\, e^{-2\frac{v_F^-}{v_F^+} \sin(\phi/2) \frac{x}{x_0}} & \,\, \text{ if }\,\, x>0
\end{aligned}
\right.
\end{equation}
where $x_0= (\hbar v_F^-)/\Delta$.
The probability density is shown in Fig.~\ref{Fig:7}. We see that the probability density develops a discontinuity at $x=0$ if  $v_F^- \ne v_F^+$ while it is continuous if $v_F^-=v_F^+$.

\vspace{0.2cm}
\noindent\underline{Discussion and Conclusions}\\
While the 1-dim BdG hamiltonian of Eq.~\eqref{hamiltonian} does not admit in general bound states, introducing a position dependent Fermi velocity and gap parameter as in Eq.~\eqref{v1} opens up the possibility of having bound states. Discrete bound states are found however only in the region $E > |\Delta|$. Interestingly we find bound states in the continuum (BIC) like solutions.  It is worth pointing out that BIC like solutions are found  both in the region $ E> |\Delta|$ as well as in the region $ E\le |\Delta|$. The two classes of BIC solutions merge continuously at $E=|\Delta|$.  This can be seen explicitly by comparing the the limiting behavior of the probability density functions, respectively  $\displaystyle\lim_{\nu \to 0^+}\rho_>(x)$ and $\displaystyle\lim_{\mu \to 0^+}\rho_>(x)$.

In particular we note how the BIC current $j_<$ in the region $E\le |\Delta|$ (superconducting  phase)  merges with the BIC current $j_>$ in the region $E> |\Delta|$ (normal  phase) ensuring continuity at $E=|\Delta|$. The same can be checked graphically by noticing that the curve $\mu=0$ of Fig.~\ref{Fig:2} coincides with the curve $\nu=0$ of Fig.~\ref{Fig:3}.

Also we note how Fig.~\ref{Fig:4} shows explicitly that in the region $E\ge |\Delta|$, the probability currents of the discrete states (full dots), c.f. Eq.~\eqref{currentdiscrete} exactly superimpose the  the probability current of the BIC states (continuous line), c.f. Eq.~\eqref{currentp}.

In conclusion we have investigated the one dimensional BdG equation with  position dependent Fermi velocity $v_F(x)$ and order parameter $\Delta(x)$. We have first considered a smooth velocity distribution $v_F(x)=v_0\,\text{cosh}(\alpha x)$ and  constant order parameter $\Delta$ showing that it leads to the  creation of bound states as well as bound states in continuum,  but no scattering solutions.  Interestingly enough we found that, with the introduction of a position dependent Fermi velocity profile, while in the normal phase ($E>|\Delta|$) are present both discrete bound states as well as BIC states, in the superconducting region ($E\le|\Delta|$) there are only BIC states. Some features of quantities like the probability density have also been analysed. It has been shown that for large $n$ the probability density becomes independent of $n$. In the case of BIC the same feature is also observed as the probability density becomes independent of energy $E$ for large values of $E$. 

 We have then considered the case of combined step-like Fermi velocity and order parameter profiles. We find that a phase difference between $\Delta^{\!\!+}$ and $\Delta^{\!\!-}$ is sufficient to provide normal (non zero) reflectance in the region $E > \text{max}\left\{\lvert\Delta^{\!\!+}\rvert,\lvert\Delta^{\!\!-}\rvert\right\}$. In the region   $E < \text{min}\left\{\lvert\Delta^{\!\!+}\rvert,\lvert\Delta^{\!\!-}\rvert\right\}$ a single bound state is found (positive branch of the spectrum). Other configurations could be the object of further work.
 

\begin{acknowledgments}
One of us (P.~R.) wishes to thank INFN Sezione di Perugia for supporting a visit during which part of this work was carried out.  He would also like to thank the Physics Department of the University of Perugia for hospitality.
\end{acknowledgments}
%

\end{document}